\documentclass[namedreferences]{solarphysics}
\usepackage[hyperref,optionalrh,solaromanenum]{spr-sola-addons} 
\usepackage{graphicx}                    
\usepackage{color}                       
\usepackage{breakurl}                         

\hypersetup{
    final=true,
    pageanchor=true,
    colorlinks=true,
    breaklinks=true,
    linkcolor=blue,
    citecolor=blue,
    urlcolor=blue,
    pdfpagemode=UseNone}

\renewcommand{\vec}[1]{{\mathbfit #1}}

\newcommand{\bb}{\vec B}

\chardef\us=`\_

\newcommand{\gkm} {\,G\,km$^{-1}$}

\begin{document}

\begin{article}

\begin{opening}

\title{The problem of the height dependence of magnetic fields in sunspots}

%
\author[addressref=aff1,corref,email={hbalthasar@aip.de}]{\inits{H. B.}\fnm{Horst}~\lnm{Balthasar}}
%
\runningtitle{The height dependence of magnetic fields in sunspots}

\address[id={aff1}]{Leibniz-Institut f\"ur Astrophysik Potsdam (AIP), An der Sternwarte 16, 14482 Potsdam, Germany}
\begin{abstract}
To understand the physics of sunspots, it is important to know the properties of their
magnetic field, and especially its height stratification plays a substantial r{\^o}le.
There are mainly two methods to assess this stratification, but they yield different 
magnetic gradients in the photospheric layers. Determinations based on the different 
origin of several spectral lines and the slope of their profiles result in gradients of 
$-2$ to $-3$\gkm, or even steeper. This is similar for the total magnetic field strength and
for the vertical component of the magnetic field. 
The other option is to determine the horizontal partial derivatives of the magnetic field,
and with the condition {$\rm{div} {\bb} = 0$}, also the vertical derivative is known.
With this method, gradients of $-0.5$\gkm{} and shallower are obtained. Obviously, these 
results do not agree. If chromospheric 
spectral lines are included, only shallow gradients around $-0.5$\gkm{} are encountered.
Shallow gradients are also found from 
gyro-resonance measurements in the radio wave range 300 -- 2000\,GHz.

Some indirect methods are also considered, but they cannot clarify the total picture.
An analysis of a numerical simulation of a sunspot indicates a shallow gradient over a 
wide height range, but with slightly steeper gradients in deep layers.

Several ideas to explain the discrepancy are also discussed. With no doubts cast on 
Maxwell's equations, the first one is to look at the uncertainties of the formation 
heights of spectral lines, but a wider range of these heights would require an extension 
of the solar photosphere that is incompatible with observations and the theory of stellar 
atmospheres. Submerging and rising magnetic flux might play a r{\^o}le in the outer penumbra,
if the resolution is too low to separate them, but it is not likely that this effect 
acts also in the umbra. A quick investigation assuming a spatial small scale structure
of sunspots together with twist and writhe of individual flux tubes shows a reduction of
the measured magnetic field strength for spectral lines sensitive to a larger height range.
However, sophisticated investigations are required to prove that the explanation for the 
discrepancy lies here, and the problem of the height gradient of the magnetic field in 
sunspots is still not solved.

\end{abstract}

%
\keywords{Sunspots, Magnetic Fields, Umbra, Penumbra, Photosphere, Chromosphere}

\end{opening}

%
\section{Introduction}\label{s:intro}

More than hundred years ago, it was proven that sunspots are 
concentrations of strong magnetic fields with a central field strength 
of 3000\,G and more \citep{Hale08}. Sunspots exist for several days or even 
weeks, and after 
the formation process, they often exhibit a stable appearance before they
finally decay. This means, sunspots are in hydrostatic equilibrium with 
their surroundings which are stratified by the gas pressure. The 
magnetic field in the sunspots causes another pressure term reducing the 
inner gas pressure. Hydrostatic equilibrium then requires that the magnetic 
field strength is height dependent and decreases with height. Many attempts 
were carried out to determine this height dependence, but up to now, 
various methods led to different height gradients. Observations based on 
different spectral lines originating in different atmospheric layers yield 
a steep gradient of the magnetic field with height ($-2$ to $-3$\gkm), while 
investigations under the divergence-free assumption yield shallower gradients 
($-0.5$\gkm). Gradients between 0 
and $-1$\gkm{} are considered as shallow and gradients beyond $-1.5$\gkm{} as steep. 
Negative values for 
these gradients indicate a decrease with height, and this convention will be 
kept throughout this review, although some authors follow the opposite convention.
Furthermore, I shall give all gradients in [\gkm], however, it is clear 
that such values are only valid for a narrow height range.
The discrepancy between the methods was 
discussed before by \cite{Solanki03} in his review on sunspots, and he states 
that no satisfactory solution was found for the small gradients in the second 
case. Nevertheless, it may be as well that the large gradients based on 
different spectral lines do not reflect the real situation on the Sun. 
So far, it is still an open question what causes these differences.

It is the main aim of this article to give an overview about the existing
observations and to discuss possible problems of the applied methods, and 
which ideas have been followed to explain the discrepancy. Section~\ref{S:deltaB} provides 
an overview about previous works based on spectral lines and their profiles,
and in Section~\ref{S:divB} investigations based on div~$\bb = 0$ are presented. Gradients in
pores are summarized in Section~\ref{S:pores}. Coronal observations at radio wavelengths 
are considered in Section~\ref{S:radio}. 
Indirect methods based on the magnetic flux and the pressure balance are presented in Sections~\ref{S:flux}
and \ref{S:p_balance}.
Numerical simulations are the topic of Section~\ref{S:numsim}. 
Finally, in the discussion (Section~\ref{S:discussion}), I consider several other possibilities 
that have been proposed to explain the discrepancy between the methods, although none of these 
explanations is a convincing solution for the problem.

\section{Observations based on different spectral lines and line profiles}\label{S:deltaB}

Spectral lines are formed in a certain range of the solar atmosphere. 
Depending on temperature, the abundance of the ion and the parameters of 
the atomic transition, this range can be quite narrow or rather extended.
In general, strong lines cover a wider range of the atmosphere and deliver
information about higher layers. Weak lines are formed in a narrow range 
which is normally located in the lower atmosphere. Observing several lines 
simultaneously allows us to probe different heights of the solar atmosphere.
In the classical way, different lines are investigated separately, 
and the magnetic gradient is calculated from the differences according to 
Equation~\ref{Eq_def_grad}. 

\begin{equation}\label{Eq_def_grad}
\frac {\Delta B} {\Delta h} = \frac {B\rm{(line~a)} - B\rm{(line~b)}}  
   {h\rm{(line~a)} - h\rm{(line~b)}}.
   \end{equation}

Nowadays, 
modern inversion techniques allow us also to evaluate several spectral lines 
simultaneously and obtain the magnetic gradient making use of the coverage 
of a wide height range. 

\subsection{Height dependence of the total and longitudinal magnetic field strength}

In this first step, publications where the height dependence of 
the total magnetic field strength or its longitudinal component was determined, are considered. 
The results are valid for the umbra if nothing else is explicitly mentioned. 
\cite{Wittmann74} investigated the spectral lines Fe\,{\sc i}\,525.02\,nm,
Fe\,{\sc i}\,617.34\,nm, Fe\,{\sc i}\,621.34\,nm, Fe\,{\sc i}\,630.25\,nm,
Fe\,{\sc i}\,633.68\,nm, and V\,{\sc i}\,625.69\,nm
obtained from spectro-polarimetric observations. He found a gradient of the magnetic 
field strength of $-1.5$\gkm. 
\cite{Pahlke90} recorded many spectral lines in the range 611\,--\,618\,nm simultaneously in
circular polarization. They used a combination of two different model atmospheres, a hot 
and a cool one, and created diagnostic diagrams for the spectral lines. The best fit was
obtained for a gradient of $-2$\gkm. An even steeper 
gradient of $-2$ to $-3$\gkm{} was found by \cite{BandS93} from the superposed 
pattern of 
the Zeeman components of three different iron lines Fe\,{\sc ii}\,614.9\,nm, 
Fe\,{\sc i}\,630.25\,nm and Fe\,{\sc i}\,684.2\,nm. For this purpose, they 
compared the observed line profiles with a set of calculated ones and selected 
that with the highest correlation coefficient.

In the last decade of the twentieth century, spectro-polarimetric inversion codes came up, and sophisticated 
polarimeters were built. To determine the magnetic gradient, there are two options.
Such inversion codes use stratified atmospheres and are thus able to determine height dependencies from 
different parts of the Stokes-profiles, even from a single line. 
The full calculation for all atmospheric layers is time-consuming. To save computing time, many codes 
perform the full calculation only for a small number of atmospheric levels (``nodes'') and 
interpolate for the levels between the nodes.
However, one has to be careful that gradients determined this way are realistic. 
Depending on where the nodes are placed, it might happen that layers with low contribution to the line 
profile influence the result and lead to unrealistic inverse gradients or ``ringing'' effects.
The other option is to invert 
profiles from two or more spectral lines originating in different atmospheric layers
independently, but assuming the magnetic parameters are height independent. The magnetic field strength 
is different for the individual lines, and from this difference  divided by the height difference,
the magnetic gradient is determined, as described in Equation~\ref{Eq_def_grad}. 
Both methods yield similar gradients, as shown in the following.

%
 \begin{figure} 
 \centerline{\includegraphics[width=\textwidth,clip=]{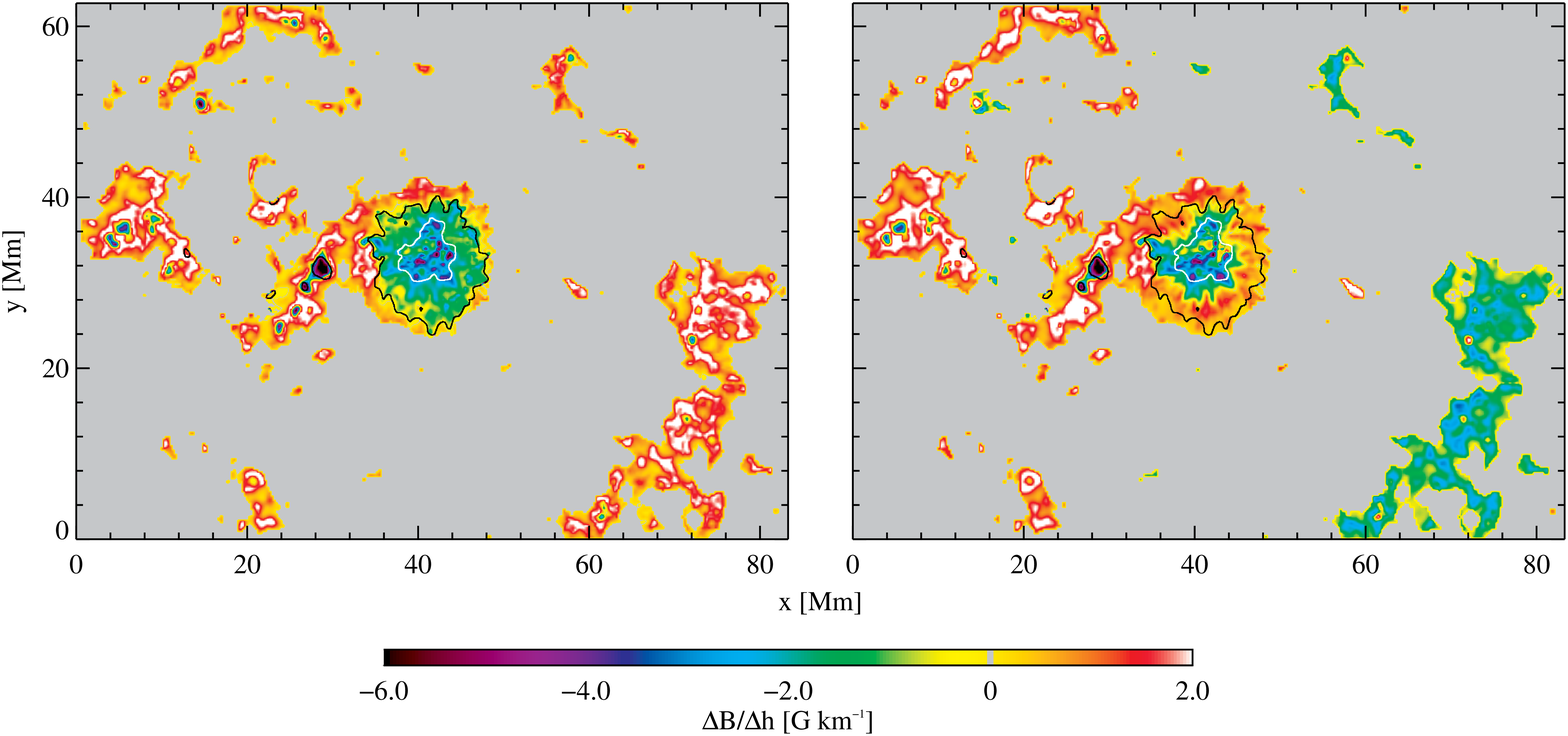}}
 \caption{Gradient of the magnetic field strength derived by \cite{BandG08} from the 
 difference of two spectral lines. \textit{Left:} total magnetic field strength. \textit{Right:} Vertical 
 component $B_{\rm z}$. Extreme values beyond $-6$\gkm{} are clipped.
 }
 \label{fig:BandG_bbh}
 \end{figure}

\cite{Collados94} used data obtained at the \textit{Gregory-Coud\'e-Telescope} (GCT) in 
Tenerife and applied the Stokes Inversion based on Response functions (SIR) code 
\citep{SIR} to create model atmospheres for different sunspots. The spectral lines 
Fe\,{\sc i}\,629.8\,nm, Fe\,{\sc i}\,630.15\,nm and Fe\,{\sc i}\,630.25\,nm were inverted 
together height-dependently for the magnetic field. For the mid 
photosphere they found a gradient of $-2.1$\gkm{} for a small ``hot'' sunspot (with an umbral temperature
of 5030\,K at lg\,$\tau = 0$) and $-0.25$\gkm{} 
for a large cool one (T = 3940\,K). They found indications that the gradients become steeper and more 
similar for different spots in deep layers of the atmosphere.
\cite{Westendorp01} analyzed data obtained at the \textit{Dunn Solar Telescope}, 
(DST, \citealp{DST}) also with the SIR code, and from the two lines 
Fe\,{\sc i}~630.15\,nm and Fe\,{\sc i}~630.25\,nm they found a gradient of 
$-1.5$ to $-2$\gkm. A similar gradient was found by \cite{Cuberes05} from a pair 
of near infrared lines, Fe\,{\sc i}\,1078.3\,nm and Si\,{\sc i}\,1078.6\,nm 
observed at the \textit{Vacuum Tower Telescope} (VTT, \citealp{VTT}) with the Tenerife
Infrared Polarimeter (TIP, \citealp{TIP}). They also used the SIR-code
and performed an inversion with a height dependent magnetic field for both lines together.
\cite{Mathew03} observed also with TIP at the VTT and used the Stokes-Profile-INversion-O-Routines code 
(SPINOR, \citealp{SPINOR}). 
They obtained even $-4.0$\gkm{} from 
the lines Fe\,{\sc i}~1564.8\,nm and Fe\,{\sc i}~1565.2\,nm which probe the 
deepest layers of the photosphere, together with two OH molecular lines recorded in 
the same spectral range. The molecular lines originate in the upper photosphere. 
In their inversion, they used six nodes at fixed optical depths for the height dependence 
of the magnetic field. 

\cite{BandG08} observed a small sunspot with TIP at the VTT and used the same spectral lines 
as \cite{Cuberes05}. Both lines are Zeeman triplets with a splitting factor of 1.5,
but they differ in the formation height. The silicon line originates higher, except for cool
umbrae. 
The two lines were recorded strictly simultaneously. Because of the small distance of only 
0.3\,nm, they fit on the same detector.
Inversions with the SIR-code were performed independently for the two lines, but without height 
dependence of the single lines.
In the umbra, they found an average gradient of $-2.6$\gkm{} and 
in the penumbra a range of $-1$ to $-2$\gkm{}. The increase of the magnetic field outside 
the penumbra is explained by the magnetic canopy. The gradient varies significantly within the umbra 
indicating an internal structure of the umbra (see Figure\,\ref{fig:BandG_bbh}).

Observations with two telescopes at the \textit{Observatorio del Teide}, Tenerife, 
were combined  by \cite{BandB09}. They observed at the VTT the same lines as \citep{BandG08}
and Fe\,{\sc i}\,630.15\,nm, Fe\,{\sc i}\,630.25\,nm and Cr\,{\sc i}\,578.2\,nm
at the \textit{T\'elescope H\'eliographique pour l'Etude du Magnetisme et des 
Instabilit\'es Solaires} (THEMIS, \citealp{gelly07}) and applied the SIR-code. They 
found a gradient of the magnetic 
field strength of up to $-3$\gkm{} in the umbra and $-2$\gkm{} for the inner penumbra.

\begin{figure} 
 \centerline{\includegraphics[width=0.98\textwidth]{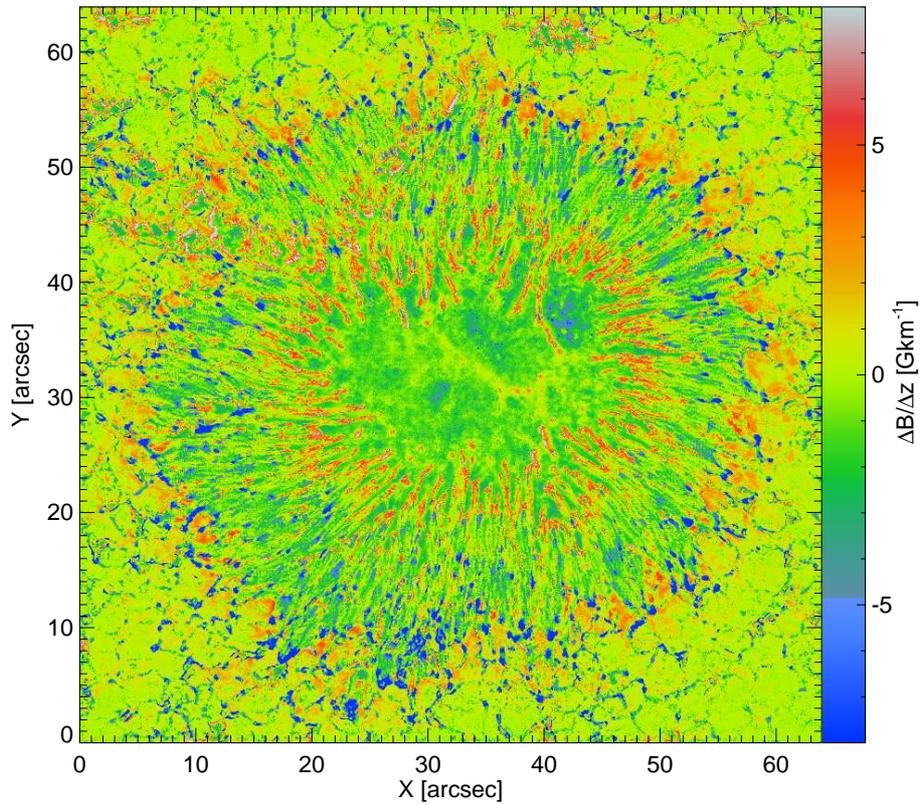}}
 \caption{Gradient of the magnetic field computed between the geometrical height 
 corresponding to the deepest nodes (log$\tau$ = $-0.9,0$). A positive field gradient 
 represents field strength increasing with height. Credit: Tiwari et al. 
 2015, 
 reproduced with permission \copyright ESO)
 }
 \label{fig:tiwari8}
 \end{figure}

\cite{Bommier13} also used THEMIS and recorded full-Stokes spectra of the lines
Fe\,{\sc i}\,630.15\,nm and Fe\,{\sc i}\,630.25\,nm. 
She inverted the line profiles separately with the UNNOFIT-code \citep{UNNOFIT} 
and found $-3$ to $-4$\gkm{} for the vertical 
gradient of the magnetic field. \cite{Betal14a} observed the same lines as \cite{BandG08}, 
inverted them with the SIR-code and obtained a gradient of $-2 \pm 0.5$\gkm{} for the
main umbra of a $\delta$-spot ($\delta$-spots harbor umbrae of both magnetic polarities 
within the same penumbra) and $-4.5 \pm 1.4$\gkm in the $\delta$-umbra. 
\cite{Tiwari15} investigated data with the same spectral lines from the Japanese {\it Hinode} spacecraft
\citep{Hinode} taken with the SOT spectropolarimeter \citep{SOT, SOT-SP}. As \cite{Mathew03}, 
they used the SPINOR-code for the inversion. In the umbra,
they found a gradient of the total magnetic field strength of $-1.5$\gkm.
In the penumbra, they obtained a very structured distribution, and in certain locations
of the inner penumbra, they even found positive gradients, as shown in 
Figure~\ref{fig:tiwari8}. \cite{Joshi2017b} confirmed such positive gradients for the
inner penumbra from photospheric lines recorded in the near infrared at the VTT and 
inverted with SPINOR.

\cite{Felipe16} found a decrease for the outer (faint) umbra 
from observations with the GREGOR solar telescope \citep{GREGOR} and SIR-inversions. For the 
central umbra and for the penumbra they found very small gradients. The spot
investigated by them exhibited a light bridge (LB) with a small magnetic field strength
at low photospheric level increasing almost to the field strength of the faint umbra at lg $\tau$ = $-3$.
Comparably small gradients of about $-1.06$\gkm{} for the umbra are reported by \cite{Verma18}, who used the 
spectral lines Si\,{\sc i}\,1082.7\,nm and Ca\,{\sc i}\,1083.9\,nm observed at the GREGOR telescope.
The same authors determined $-0.69$\gkm{} for the normal penumbra. This spot exhibited an elongated umbral
core (EUC) embedded in an irregular part of the penumbra, and here the gradient was $-1.31$\gkm.

Very sensitive to the magnetic field is the magnesium line at 12.32\,$\mu$m, because 
the sensitivity scales with the wavelength. On the other hand, the spatial 
resolution decreases, and more efforts (cooling) are needed to operate detectors
for such wavelengths. Therefore, only a few attempts have been undertaken to
used this line to determine the height dependence of the magnetic 
field.  \cite{Bruls95} compared a series of synthetic line profile with observations of
\cite{Hewagama93} at the \textit{McMath-Pierce telescope} \citep{McMath} at the \textit{Kitt Peak Observatory}
and found a range of $-0.7$ to $-3$\gkm{} for the gradient in 
penumbrae, the steeper values occur close to the umbra.
\cite{Moran2000} compared this line with Fe\,{\sc i}~1564.8\,nm with Fe\,{\sc i}\,1564.8\,nm
and reported a gradient of $-2$\gkm{} for the umbra and $-1$\gkm{} for the penumbra.

%
 \begin{table}
 \caption{Magnetic field gradients from the difference of photospheric spectral lines.}
 \label{tbl:phot}
 \begin{tabular}{lccc}     
 \hline
  Author                 &  spectral lines         &  umbra             &  penumbra \\
  \hline
  \cite{Wittmann74}      & several lines           & $-1.5$\gkm         & \rule[-2.5mm]{0mm}{3mm}\\
  Pahlke and Wiehr       & several lines           & $-2.0$\gkm         & \\ 
   (1990)                &                         &                    & \rule[-2.5mm]{0mm}{3mm}\\ 
  Balthasar and          & Fe\,{\sc i}\,684.2\,nm  & $-2$ -- $-3$\gkm   & \\
  Schmidt (1993)         & Fe\,{\sc i}\,630.25\,nm &                    & \\
                         & Fe\,{\sc ii}\,614.9\,nm &                    & \rule[-2.5mm]{0mm}{3mm}\\                       
  \cite{Collados94}      & Fe\,{\sc i}\,629.78\,nm & $-2.1$\gkm         & \\
                         & Fe\,{\sc i}\,630.15\,nm & $-0.25$\gkm (dark U) & \\
                         & Fe\,{\sc i}\,630.25\,nm &                    & \rule[-2.5mm]{0mm}{3mm} \\  
  \cite{Bruls95}         & Mg\,{\sc i}\,12.32\,$\mu$m &                 & $-0.7$ -- $-3$\gkm  \rule[-2.5mm]{0mm}{3mm} \\
  \cite{Moran2000}       & Mg\,{\sc i}\,12.32\,$\mu$m &  $-2$\gkm       & $-1$\gkm   \\
                         & Fe\,{\sc i}\,1564.8\,nm &                    &  \rule[-2.5mm]{0mm}{3mm}\\
  Westendorp Plaza       & Fe\,{\sc i}\,630.15\,nm & $-1.5$ -- $-2$\gkm   &  \\ 
  {\it et al.} (2001)    & Fe\,{\sc i}\,630.25\,nm &                    & \rule[-2.5mm]{0mm}{3mm}\\
  \cite{Mathew03}        & Fe\,{\sc i}\,1564.8\,nm & $-4.0$\gkm         & \\  
                         & Fe\,{\sc i}\,1565.2\,nm &                    & \\
                         & OH\,1565.19\,nm         &                    & \\
                         & OH\,1565.35\,nm         &                    & \rule[-2.5mm]{0mm}{3mm}\\
  S{\'a}nchez Cuberes    & Fe\,{\sc i}\,1078.3\,nm &                    & \\
  {\it et al.}  (2005)   & Si\,{\sc i}\,1078.6\,nm &                    & \rule[-2.5mm]{0mm}{3mm}\\
  Balthasar and          & Fe\,{\sc i}\,1078.3\,nm &  $-2.6$\gkm        & $-1$ -- $-2$\gkm \\
   G\"om\"ory (2008)     & Si\,{\sc i}\,1078.6\,nm &                    & \rule[-2.5mm]{0mm}{3mm}\\
  Balthasar and          & several lines           & $-3$\gkm           &  $-2$\gkm \\
   Bommier (2009)        &                         &                    & \rule[-2.5mm]{0mm}{3mm}\\
  \cite{Bommier13}       & Fe\,{\sc i}\,630.15\,nm & $-3$ -- $-4$\gkm   & \\
                         & Fe\,{\sc i}\,630.25\,nm &                    & \rule[-2.5mm]{0mm}{3mm}\\
  Balthasar {\it et al.} & Fe\,{\sc i}\,1078.3\,nm & $-2.0$\gkm         & \\
   (2014a)               & Si\,{\sc i}\,1078.6\,nm & $-4.5$\gkm ($\delta$-U)  & \rule[-2.5mm]{0mm}{3mm}\\
  \cite{Tiwari15}        & Fe\,{\sc i}\,630.15\,nm & $-0.5$\gkm         & $< 0$  outer PU \\                        
                         & Fe\,{\sc i}\,630.25\,nm & $-1.5$\gkm         & $> 0$ inner PU \rule[-2.5mm]{0mm}{3mm}\\
  \cite{Felipe16}        & Si\,{\sc i}\,1082.7\,nm & $< 0$              & $\approx 0$  \\
                         & Ca\,{\sc i}\,1083.9\,nm & $> 0$ (LB)         & \rule[-2.5mm]{0mm}{3mm}\\
  \cite{Joshi2017a}      & Si\,{\sc i}\,1082.7\,nm & $< 0$              & $> 0$ inner PU\\
                         & Ca\,{\sc i}\,1083.3\,nm &                    &  \rule[-2.5mm]{0mm}{3mm}\\ 
  \cite{Verma18}         & Si\,{\sc i}\,1082.7\,nm & $-1.1$\gkm         & $-0.7$\gkm \\
                         & Ca\,{\sc i}\,1083.9\,nm &                    & $-1.3$\gkm (EUC) \rule[-2.5mm]{0mm}{3mm}\\                      
        \hline
 \end{tabular}
 \end{table}

All spectral lines discussed so far originate in the photosphere. In general it
is more difficult to determine the magnetic field strength from  
chromospheric lines. There is only a limited selection of spectral lines coming from
the chromosphere, most of them belong to light and abundant ions, thus the components 
of the line profiles are broader. Local thermal equilibrium (LTE) is no longer valid for 
these lines, and a proper inversion requires more efforts than for photospheric lines.
In addition, most of these lines are not Zeeman triplets,
and their effective Land{\'e} factor is two or even less. Nevertheless, chromospheric
lines have been used to determine chromospheric magnetic fields, but often only with 
approximations to estimate the magnetic field.
Results obtained in combination with photospheric lines obtained by \cite{Abdussamatov71}, 
\cite{Ruedi95}, \cite{Liu96}, \cite{Orozco05}, \cite{Berlicki06}, 
\cite{Schad15}, \cite{Joshi16}, and \cite{Joshi2017a} are summarized in 
Table\,\ref{tbl:chrom}. 
\cite{Abdussamatov71} recorded the spectral lines H$\alpha$, Na\,D$_1$ and D$_2$ and 
Fe\,{\sc i}\,630.25\,nm on photographic plates and found gradients $-0.6$\gkm{} for strong magnetic fields 
($>2000$\,G) and $-0.8$\gkm{} for weaker fields. \cite{Ruedi95} measured the circular polarization of the 
helium lines around 1083\,nm with the 
main spectrograph of the \textit{McMath-Pierce telescope}
and compared them 
with observations in the line Fe\,{\sc i}\,1089.6\,nm with the \textit{Fourier-Transform-Spectrometer}. 
They determined gradients between $-0.38$ and $-0.5$\gkm{} for the umbra and shallower gradients for the penumbra.
\cite{Liu96} observed the H$\beta$ line and Fe\,{\sc i}\,532.4\,nm with the \textit{Solar Magnetic Field Telescope} 
at the \textit{Huairou Solar Observing Station}, where they obtained photospheric vector magnetograms and 
chromospheric longitudinal magnetograms. They determined a range of $-0.7$ to $-0.8$\gkm{} for 
the magnetic gradient
from the difference method.  \cite{Orozco05} performed simultaneous observations at the VTT with TIP in 
the lines He\,{\sc i}\,1083\,nm and Si\,{\sc i}\,1082.7\,nm. They inverted the silicon line with SPINOR and the 
helium line with the H{\sc e}LI{\sc x}-code \citep{HELIX}. Assuming a height difference of 2000\,km 
between the two lines, they found gradients between $-0.5$ and $-1.0$\gkm.
\cite{Berlicki06} measured the line-of-sight (LOS)-component of
the magnetic field at THEMIS and obtained a relative gradient of $-6.2 \times 10^{-4}$\,km$^{-1}$.
Depending on the reference value, this amounts to $-0.6$ to $-1.6$\gkm.
\cite{Schad15} used the \textit{Facility Infrared Spectropolarimeter} (FIRS, \citealp{FIRS}) at the DST
to record the helium and the silicon line at 1083\,nm. Their result of $-0.5$\gkm{} is in the same order 
as comparable investigations. \cite{Joshi2017a} observed with TIP at the VTT and found also values between 
$-0.5$ and $-0.9$\gkm.

%
 \begin{table}
 \caption{Magnetic field gradients from the combination of chromospheric 
  and photospheric spectral lines.}\label{tbl:chrom}
 \begin{tabular}{lccc}     
 \hline
  Author                       &  spectral lines         &  umbra               &  penumbra \\
  \hline
  \cite{Abdussamatov71}        & H$\alpha$               & $-0.6$ -- $-0.8$\gkm & \\
                               & Na~D$_1$                &                      & \\
                               & Fe\,{\sc i}\,630.25\,nm &                      & \rule[-2.5mm]{0mm}{3mm}\\
  \cite{Henze82}               & C\,{\sc iv}\,154.6\,nm  & $-0.4$ -- $-0.6$\gkm & \\ 
                               & Fe\,{\sc i}\,525.0\,nm  &                      & \rule[-2.5mm]{0mm}{3mm}\\
  \cite{Hagyard83}             & C\,{\sc iv}\,154.6\,nm  & $-0.36$\gkm          & \\
                               & Fe\,{\sc i}\,525.0\,nm  &                      & \rule[-2.5mm]{0mm}{3mm}\\
  R\"uedi {\it et al.} (1995)  & He\,{\sc i}\,1083.0\,nm & $-0.4$ -- $-0.6$\gkm & $-0.1$ -- $-0.3$\gkm \\ 
                               & Fe\,{\sc i}\,1089.6\,nm &                      & \\
                               & Fe\,{\sc i}\,1078.3\,nm &                      & \rule[-2.5mm]{0mm}{3mm}\\
  \cite{Liu96}                 & H$\beta$                & $-0.7$ -- $-0.8$\gkm & \\
                               & Fe\,{\sc i}\,532.42\,nm &                      & \rule[-2.5mm]{0mm}{3mm}\\
  Orozco~Su\'arez {\it et al.} & He\,{\sc i}\,1083.0\,nm & $-0.5$ -- $-1.0$\gkm & \\
     (2005)                    & Si\,{\sc i}\,1082.7\,nm &                      & \rule[-2.5mm]{0mm}{3mm}\\
  Berlicki {\it et al.} (2006) & Na~D$_1$                & $-0.6$ -- $-1.6$\gkm & \\
                               & Ni\,{\sc i}\,676.8\,nm  &                      & \rule[-2.5mm]{0mm}{3mm}\\ 
  \cite{Schad15}               & He\,{\sc i}\,1083.0\,nm & $-0.5$\gkm           & \\                        
                               & Si\,{\sc i}\,1082.7\,nm &                      & \rule[-2.5mm]{0mm}{3mm}\\
  \cite{Joshi16}               & He\,{\sc i}\,1083.0\,nm &                      & $-0.6$\gkm{} inner PU \\
                               & Si\,{\sc i}\,1082.7\,nm &                      & $-0.2$\gkm{} outer PU \rule[-2.5mm]{0mm}{3mm}\\
  \cite{Joshi2017a}            & He\,{\sc i}\,1083.0\,nm & $-0.5$ -- $-0.9$\gkm & $-0.1$ -- $-0.4$\gkm \\
                               & Si\,{\sc i}\,1082.7\,nm &                      & outer to inner PU \\
                               & Ca\,{\sc i}\,1083.3\,nm &                      &  \rule[-2.5mm]{0mm}{3mm}\\ 
  \hline
 \end{tabular}
 \end{table}

\cite{Henze82} and \cite{Hagyard83} combined ground based observations with 
space based recordings of a C\,{\sc iv} line at 154.8\,nm with the 
\textit{Ultraviolet Spectrometer and Polarimeter}
(UVSP, \citealp{UVSP}) on board the \textit{Solar Maximum} mission. Their results are also listed in 
Table\,\ref{tbl:chrom}, but they do not differ significantly from only ground based observations. 
All these values are smaller than those from investigations 
where only photospheric lines were used. The height difference $\Delta h$ is 
1000\,km or more, while it is of the order of 100 -- 200\,km in the photospheric 
case. 
For the penumbra, the gradients are somewhat shallower than for the umbra, especially in the outer penumbra. 
A recent example is given by \cite{Joshi16},
who used GRIS at GREGOR to record the lines He\,{\sc i}\,1083.0\,nm and Si\,{\sc i}\,1082.7\,nm.

\subsection{Height dependence of the vertical component of the magnetic field}  

If the full Stokes vector is measured, and the polarimetric azimuth ambiguity is solved, it becomes possible to 
investigate the height dependence of the vertical component of the magnetic field $B_z$. Here one has 
to keep in mind that the sign of $B_z$ can be negative in contrast to the total field strength. Then a 
decrease of the total field strength coincides with a positive gradient of the vertical 
magnetic component. In the following, I ignore the sign of $B_z$ and show a negative gradient when 
the absolute values of the vertical component of the magnetic field strength decreases with height. 
However, the absolute value
of the vertical component may increase with height, which is often the case in the outer penumbra. 
Results of such investigations are listed in 
Table\,\ref{tbl:b_z}. Some of these results are based on the same data sets discussed already in 
the previous section.

%
 \begin{table}
 \caption{Gradients of the vertical magnetic field component $B_z$.}
 \label{tbl:b_z}
 \begin{tabular}{lccc}     
 \hline
  Authors             &  spectral lines         &  umbra                &  penumbra \\
 \hline 
  \cite{Eibe02}       & Na~D$_1$                & $-0.2$ -- $-1.0$\gkm  & \rule[-2.5mm]{0mm}{3mm}\\           
  Balthasar and G\"om\"ory     & Fe\,{\sc i}\,1078.3\,nm & $-2.15 \pm 0.06$\gkm  & 1--2\gkm{} \\
    (2008)    (VTT)   & Si\,{\sc i}\,1078.6\,nm &                       & (outer PU) \rule[-2.5mm]{0mm}{3mm}\\
  \cite{Betal13}      & Fe\,{\sc i}\,1078.3\,nm & $-1$ -- $-2$\gkm      &\\
         (VTT)        & Si\,{\sc i}\,1078.6\,nm &                       &\\
   ({\it Hinode})     & Fe\,{\sc i}\,630.15\,nm & $-2$ -- $-3$\gkm      &\\
                      & Fe\,{\sc i}\,630.25\,nm &                       &\rule[-2.5mm]{0mm}{3mm}\\   
  \cite{Bommier13}    & Fe\,{\sc i}\,630.15\,nm & $-3$ -- $-4$\gkm      &\\
      (THEMIS)        & Fe\,{\sc i}\,630.25\,nm &                       &\rule[-2.5mm]{0mm}{3mm}\\
  \cite{Betal14b}     & Fe\,{\sc i}\,1078.3\,nm & $-2$ -- $-4$\gkm      &\\
         (VTT)        & Si\,{\sc i}\,1078.6\,nm & $-6$\gkm ($\delta$-um.)  &\rule[-2.5mm]{0mm}{3mm}\\
  \cite{Jaeggli12b}   & Fe\,{\sc i}\,630.2\,nm  & 0 -- $-3$\gkm         & $-5$--$+4$\gkm \\
                      & Fe\,{\sc i}\,1565.\,nm  &                       & \rule[-2.5mm]{0mm}{3mm}\\
  \cite{Verma18}      & Ca\,{\sc i}\,1083.9\,nm & $-0.96$\gkm           & $-0.09$\gkm\\
  (GREGOR GRIS)       & Si\,{\sc i}\,1078.6\,nm & $-1.12$\gkm (EUC)     &\rule[-2.5mm]{0mm}{3mm}\\
   \hline
 \end{tabular}
 \end{table}

The results of \cite{BandG08} are displayed in the right panel of 
Figure\,\ref{fig:BandG_bbh}. Similar values were obtained by \cite{Betal13}. \cite{Bommier13}
found steep gradients from THEMIS observations. For the special case of a $\delta$-spot, 
\cite{Betal14b} also found steep gradients, especially for the $\delta$-umbra.
\cite{Eibe02} investigated the LOS-component
of the magnetic field determined from the Na\,D$_1$ line. The sunspot observed by them
was located about 20$^\circ$ off the disk center, thus the LOS-component is an approximation
for the vertical component of the magnetic field. They determined the decrease relative
to the value of the measured field strength and obtain 1.$\times 10^{-3}$\,km$^{-1}$.
They measured field strengths up to 1000\,G, and with these values, one gets the 
values given in Table\,\ref{tbl:b_z}. This investigation of \cite{Eibe02} is the only one 
considering a single chromospheric line listed in Table\,\ref{tbl:b_z}. Therefore the values 
are clearly lower than the others, where also photospheric lines were included. 
While the investigation of \cite{Jaeggli12b} shows a wide variation from $-5$ to $+4$\gkm, 
\cite{Verma18} found shallower 
gradients around $-1$\gkm. \cite{Jaeggli12b} compared simultaneous observations with the FIRS 
at the DST of the lines Fe\,{\sc i}\,1565\,nm and
Fe\,{\sc i}\,630.25\,nm, and inverted them with the Milne-Eddington code Two-Component-Magneto-Optical code
(2CMO) based on a formalism of \cite{Jefferies89}. 

Summarizing, in the umbra, similar gradients are found as for the total magnetic field strength. They are 
slightly shallower
since the absolute value of $B_z$ in general is smaller than the total field strength. In the outer penumbra, 
the opposite sign is found in several cases. The magnetic field in deep layers is almost horizontal. Thus
the vertical component is very small and increases to higher layers where the magnetic field is less inclined.

\section{Observations based on divergence-free condition}\label{S:divB}

%
 \begin{figure} 
 \centerline{\includegraphics[width=\textwidth, clip=]{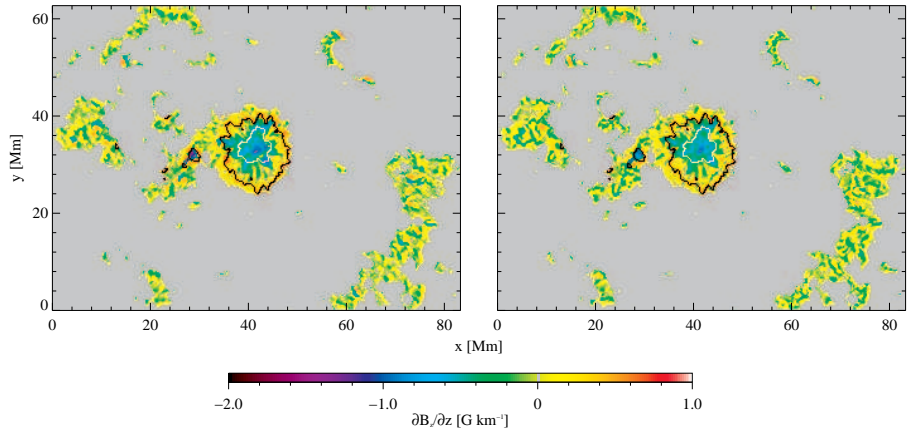}}
 \caption{Gradient of the vertical component of the magnetic field strength following the 
 divergence-free condition taken from the results of \cite{BandG08}. \textit{Left:} for 
 Fe\,{\sc i}\,1078.3\,nm. \textit{Right:} for Si\,{\sc i}\,1078.6\,nm. 
 Extreme values beyond $-2$\gkm{} are clipped.}
 \label{fig:dbzdz}
 \end{figure}

Maxwell's equations require that the magnetic field is divergence-free (or source-free, see 
Equation~\ref{Eq_divB}).

\begin{equation}\label{Eq_divB}
\rm{div} {\bb} = 0 =\frac {\partial B_{\rm x}}{\partial x} + \frac {\partial B_{\rm y}}
{\partial y} + \frac{\partial B_{\rm z}}{\partial z}.  
\end{equation}

This equation allows us to determine the partial derivative of the vertical 
component of the magnetic field $B_{\rm z}$ from the horizontal variation of 
the magnetic field, which can be observed by using two-dimensional 
spectropolarimeters or by scanning with a classical spectrograph. The partial derivatives 
are approximated by quotients of the differences between neighboring pixels.

\begin{equation}\label{Eq_dbzdz}
 \frac{\partial B_{\rm z}}{\partial z} \approx -\frac{\Delta B_{\rm x}}{\Delta x}
- \frac{\Delta B_{\rm y}}{\Delta y}.
\end{equation}

This method has been applied by \cite{Hagyard83}, \cite{Hofmann89}, \cite{Liu96}, 
\cite{Balthasar06},
\cite{BandG08}, and \cite{Betal13} to different photospheric spectral lines. 
The data sets of most of these authors are the same as described in the previous sections.
\cite{Hofmann89} used data obtained in the line Fe\,{\sc i}\,525.3\,nm at the \textit{Einstein-tower} in Potsdam
\citep{Einsteinturm}
and in Fe\,{\sc i}\,525.0\,nm at the \textit{Sayan Observatory} near Irkutsk \citep{Sayan}.
The results for the partial derivative of the vertical component of the magnetic field
are typically in the order of $-0.5$\gkm{} and are summarized in Table~\ref{tbl:div0}. 
These values are much
shallower than those from the difference method, even when the same data sets 
are used. The results from this method compare more to those obtained from 
photospheric - chromospheric comparisons, see Table\,\ref{tbl:chrom}.

As an example, the results of \cite{BandG08} are displayed in Figure\,\ref{fig:dbzdz}.
\cite{Bommier13} found values of the order of $-0.4$ to $-0.5$\gkm.
\cite{Betal14b} obtained the main umbra of a $\delta$-spot a shallow gradient of $-0.5$\gkm for $B_{\rm z}$ 
($B_{\rm z}$ is negative, therefore in Figure\,9 of \citealp{Betal14b} the gradient 
appears positive)
and a much steeper gradient of $-2.0$\gkm in the $\delta$-umbra of positive polarity.
There is a weak trend that more recent data with higher spatial resolution
yield steeper gradients of the vertical magnetic field with height.

%
 \begin{table}
 \caption{Gradients of the (absolute) vertical magnetic field component from the 
 divergence-\-free condition.}
  \label{tbl:div0}
 \begin{tabular}{lccc}     
 \hline
  Author                    &  spectral lines         &  umbra                 &  penumbra \\
  \hline
  \cite{Hagyard83}          & C\,{\sc iv}\,154.6\,nm  & $-0.21$\gkm            & \\
                            & Fe\,{\sc i}\,525.0\,nm  &                        & \rule[-2.5mm]{0mm}{3mm}\\
  Hofmann and               & Fe\,{\sc i}\,525.0\,nm  & $-0.32$\gkm            & \\
   Rendtel (1989)           & Fe\,{\sc i}\,525.3\,nm  &                        & \rule[-2.5mm]{0mm}{3mm}\\
  \cite{Liu96}              & Fe\,{\sc i}\,532.42\,nm & $-0.66$ -- $-0.77$\gkm & \rule[-2.5mm]{0mm}{3mm}\\
  \cite{Balthasar06}        & Fe\,{\sc i}\,1564.8\,nm & $-0.5$ -- $-1.5$\gkm   & \\
                            & Fe\,{\sc i}\,1089.6\,nm &                        & \rule[-2.5mm]{0mm}{3mm}\\     
  Balthasar and             & Fe\,{\sc i}\,1078.3\,nm & $-0.5$ -- $-1.0$\gkm   & 0 -- $-0.5$\gkm \\
   G\"om\"ory (2008)        & Si\,{\sc i}\,1078.6\,nm &                        &\rule[-2.5mm]{0mm}{3mm} \\            
  \cite{Betal13}            & Fe\,{\sc i}\,1078.3\,nm & $-0.5$ -- $-1.0$\gkm   & 0 -- $-0.5$\gkm \\
                            & Si\,{\sc i}\,1078.6\,nm &                        & \rule[-2.5mm]{0mm}{3mm}\\ 
  \cite{Bommier13}          & Fe\,{\sc i}\,630.15\,nm & $-0.4$ -- $-0.5$\gkm   & \\
                            & Fe\,{\sc i}\,630.25\,nm &                        & \rule[-2.5mm]{0mm}{3mm}\\  
  \cite{Betal14b}           & Fe\,{\sc i}\,1078.3\,nm & $-0.5$\gkm             & \\
                            & Si\,{\sc i}\,1078.6\,nm & $-2$ \gkm ($\delta$-U) &\rule[-2.5mm]{0mm}{3mm} \\
 \hline
 \end{tabular}
 \end{table}

At the outer edge of the penumbra and just outside, gradients of opposite sign are often found,
indicating an increase of the vertical magnetic field component with height.
Inside the penumbra, this can be explained by the inclination of the magnetic field.
In deep layers, the magnetic field is almost horizontal with a very small vertical
component. In higher layers, the magnetic field is less inclined, and the vertical 
component becomes significant. Just outside the penumbra, we encounter granulation,
so that the total magnetic field is small in the deep layers. Further up, there is a 
magnetic canopy with a higher field strength.

\section{Pores}\label{S:pores}

So far, we considered mature sunspots with a penumbra, but the same methods can be applied also to
solar pores without a penumbra. Pores have a smaller magnetic field strength than sunspot umbrae,
values of about 2000\,G are found. The magnetic gradient is even steeper than for the umbra of
larger sunspots. \cite{muglach94} determined $-5$\gkm{} from a broadening excess of the $\sigma$-components 
of the Stokes-$V$ profile of the line Fe\,{\sc i}\,1564.8\,nm in comparison to Fe\,{\sc i}\,1565.2\,nm.
\cite{Suetterlin98} measured gradients of $-4$\gkm{} based on the bisectors of the Stokes-$V$ lobes. 
In a pore next to the sunspot, \cite{BandG08} found 
a value of even $-8.2$\gkm{} for the total magnetic field strength and $-5.6$\gkm{} for the vertical component 
from the difference method. This pore is visible in Figure\,\ref{fig:BandG_bbh} left of the main spot.  
From the divergence-free condition, the same authors obtained $-0.7$ to $-1.0$\gkm{} (see Figure\,\ref{fig:dbzdz}).
There is a trend that pores exhibit steeper gradients than large spots. Such a trend is in agreement 
with model calculations presented by \cite{Oshi84}.

%
 \begin{figure} 
 \centerline{\includegraphics[width=0.98\textwidth,clip=]{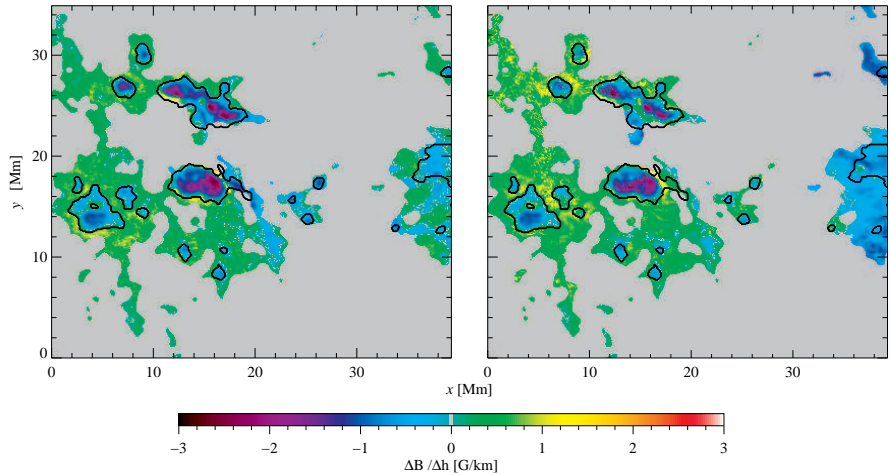}}
 \caption{Gradient of the magnetic field strength of the pores investigated by \cite{Betal16} and 
 \cite{Betal18} from the 
 difference of two spectral lines. \textit{Left:} Total magnetic field strength. \textit{Right:} Vertical 
 component $B_{\rm z}$. 
 Extreme values beyond $-3$\gkm{} are clipped.
 }
 \label{fig:pores_dbdh}
 \end{figure}

A group of pores connected via an arch filament system with a spot of opposite polarity was investigated 
by \cite{Betal16} and \cite{Betal18}. The observations were performed with the GREGOR Infrared spectrograph
(GRIS, \citealp{GRIS} at the GREGOR telescope. Only the magnetic field strength was determined
from the spectral lines Ca\,{\sc i}\,1083.9\,nm and the Si\,{\sc i}\,1082.7\,nm, but magnetic gradients
were not investigated. 
Therefore I revisited the data and present the height dependence of the magnetic 
field in the pores in Figure\,\ref{fig:pores_dbdh}. The steepest gradient of the total magnetic field strength
in the main central pore amounts to $-2.8$\gkm. The average over the larger pores is $-1.32 \pm \,0.01$\gkm.
For the vertical component of the magnetic field an average gradient of $-2.08 \pm \,0.03$\gkm{} is obtained. 
The vertical component decreases steeper with height than the total field strength indicating that the 
magnetic field becomes more horizontal at a certain layer above the pores. The steepest gradient of the 
vertical component amounts to  $-2.9$\gkm{} and appears in one of the merging pores located above the 
central pore in 
Figure\,\ref{fig:pores_dbdh}. (For the temporal evolution of the pores see 
\citealp{Betal16}).
 
Due to the short exposure times, these data are noisy, and the determination of magnetic gradients  
according to the divergence-free condition suffers from this limitation. While the mean values inside the pores 
are $-1.0$\gkm{} for the calcium line and $-1.2$\gkm{} for the silicon line, there are lanes dividing 
the pores into two halves where gradients of $-2$\gkm{} and steeper are encountered.

\section{Radio observations}\label{S:radio}

To measure magnetic fields in the corona, radio observations provide several possibilities, for an overview see 
\cite{White05}.
Magnetic field strengths of sunspots can be determined when gyro resonance emission appears above umbrae of 
sufficiently large size. Electrons spiral along magnetic field lines and emit radio radiation at harmonics of 
the electron gyro frequency, which is $\nu = B$/357\,GHz. Thus, the magnetic field strength can be determined  
according to the equation $B = 357\nu\,n^{-1}$, where $n$ is the order of the harmonic.
However, radio telescopes with spatial resolution operate only at selected frequencies.
\cite{Akhmedov82} used the \textit{RATAN-600} radio telescope \citep{RATAN600} at five different frequencies and 
measured also the degree of circular polarization. For different sunspots, they found field strengths between 
1420\,G and 2200\,G. Together with the photospheric magnetic field strengths as they were published in 
\textit{Solnechnye Dannye}, they arrived at a magnetic gradient shallower than $-0.25$\gkm. 
In general, the photospheric field strengths were between 200 and 600\,G higher than those  
determined from the radio emission. A problem with such measurements
on the solar disk is that the height of the origin of the radio emission can only be estimated under special
assumptions and modeling.

\cite{BandW06} observed layers above a sunspot at the solar limb with the \textit{Very Large Array} (VLA)
at frequencies of 8 and 15\,GHz. From these spatially resolved observations above the solar limb, they precisely
determined the height above the photosphere.
They argued that the emission at 15\,GHz 
corresponds to the third harmonic with a magnetic field strength of 1750\,G 
at 8000\,km above the photosphere. Similarly, they conclude that the field strength is 960\,G
at 12000\,km. Assuming a photospheric 
field strength of 2600\,G, these values lead to gradients of $-0.11$\gkm{} and $-0.14$\gkm, respectively.

The atmosphere above sunspots is also investigated by \cite{Stupishin18} based on \textit{RATAN-600} observations.
In their Figure 3 they show  gyro-resonance contours superposed on magnetograms for two different heights,
but they do not explicitly discuss magnetic gradients. Taking the extreme values from both layers, one 
obtains a magnetic gradient of $-0.25$\gkm.

\section{Magnetic flux considerations}\label{S:flux}

%
 \begin{figure} 
 \centerline{\includegraphics[width=0.8\textwidth, clip=]{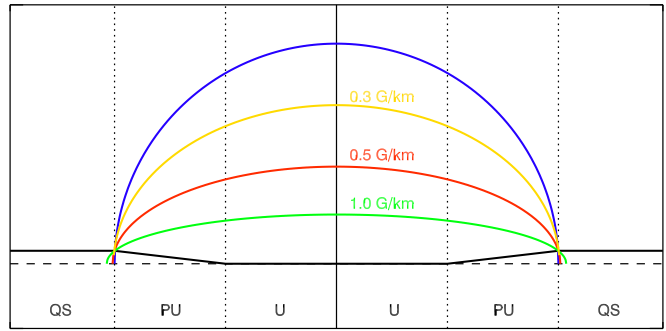}}
 \caption{Sketch to illustrate the height of ellipsoidal caps with different height gradients 
 of the magnetic field. The blue curve represents a hemisphere above the sunspot, and different colors stand 
 for different height gradients. 
 }
 \label{fig:ellipsoid}
 \end{figure}

\begin{figure} 
 \centerline{\includegraphics[height=0.8\textwidth, angle=90, clip=]{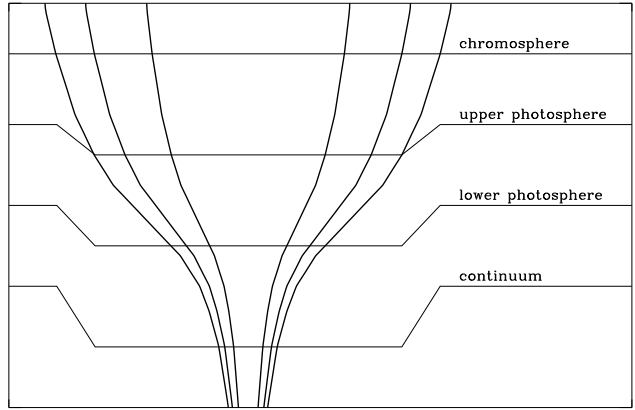}}
 \caption{Sketch of goblet-like configuration of the magnetic field.  
 }
 \label{fig:goblet}
 \end{figure}

To probe whether the penumbra is deep or shallow, \cite{SandS93} estimated the magnetic flux going through 
a dome-like hemisphere above the sunspot with a constant field strength as measured at the outer edge of 
the penumbra.
Then they measured the flux of the umbra and found this value to be much smaller than the flux going through 
the hemisphere and concluded for a deep penumbra. \cite{BandC05} turned this idea around and measured the full 
Stokes-vector for a map covering a whole sunspot. From this map they determined the magnetic flux rising 
through the photosphere and found that it is less than the amount that would leave through a hemisphere. 
The measured amount of flux can leave 
through an ellipsoidal cap with a top height of 5250\,km, while the radius of the spot is 9500\,km. A Wilson 
depression of 800\,km was assumed for these investigations. A sketch is 
shown in Figure\,\ref{fig:ellipsoid}. \cite{BandG08} confirmed this result, which is in agreement with the 
observations including a wider height range, but not with measurements from different photospheric lines.
If this picture of a flat ellipsoid with a constant magnetic field strength represents an approximation 
to the real situation in a sunspot, it means that the magnetic field is less inclined in the chromosphere 
than in the photosphere. Magnetic field lines cross such a surface perpendicularly, and within the ellipsoid, 
the field has to expand horizontally. The magnetic field 
configuration would be goblet-like as the sketch in Figure~\ref{fig:goblet} demonstrates. However, the magnetic 
field may be strongly structured in the upper chromosphere, and then such a picture will be invalid.

%

\section{Horizontal pressure balance}\label{S:p_balance}

Once mature sunspots are formed, they appear rather stable for several days. Therefore, it is expected 
that they are in horizontal pressure equilibrium with their surroundings. The horizontal pressure balance 
is represented by Equation~\ref{Eq_phor},

\begin{equation}\label{Eq_phor}
P_{\rm ge} = P_{\rm gi} + \frac {B_{\rm z}^2}{80\pi} + P_{\rm mc},
\end{equation}

\noindent
where $P_{\rm ge}$ is the external and $P_{\rm gi}$ the internal gas pressure. The magnetic pressure consists of 
two terms, where term $P_{\rm mc}$ represents the pressure due to the curvature of the field lines, which is often 
neglected because it is not easy to determine, and it is expected that this term is smaller than the 
main magnetic pressure term. 
However, this expectation is only justified when the curvature of the magnetic field lines is small enough, 
and with a larger curvature, the second term might be of the same order as the pressure term. 
The pressure is measured in Pa, therefore the factor 80.

\cite{Setal93} investigated two different scenarios and put an upper limit of the decrease of the magnetic 
field to 2\gkm{} and 0.8\gkm, respectively. The scenarios differ in the jump of the Wilson depression at the 
umbral boundary. \cite{Mathew04} determined a map of the Wilson depression based on this method, but they 
gave no statement of the magnetic field gradient. \cite{Petal10} investigated a small area in the inner 
penumbra and obtained a gradient of $-0.2$\gkm. \cite{Suetterlin98} compared a model derived from pore observations 
with a photospheric model of \cite{Vernazza76} and determined the magnetic field strength decrease to less than
5\,--\,6\gkm. An investigation for penumbral flux tubes was presented by \cite{Borrero07}, showing that the 
force equilibrium is guaranteed if the magnetic field possesses a transverse component.

%
 \begin{figure} 
 \centerline{\includegraphics[width=0.9\textwidth, clip=]{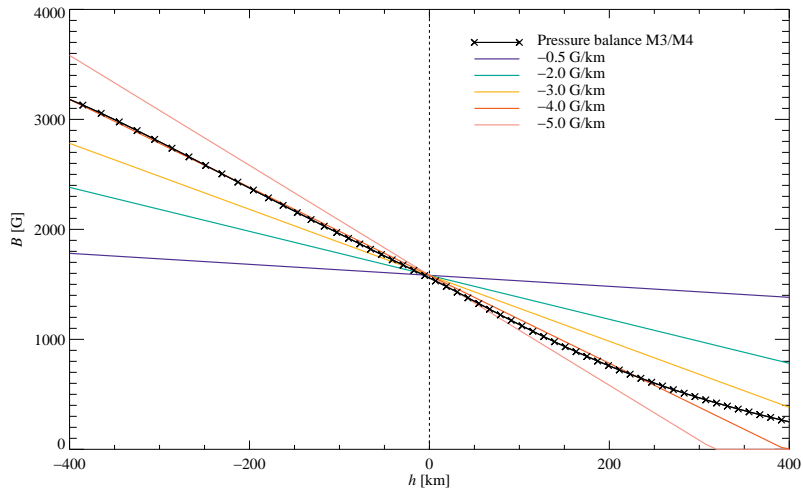}}
 \caption{Sketch of the magnetic pressure balance. The black curve represents the height dependence of the
 magnetic field strength in the central umbra if the total pressure is balanced with that of the quiet Sun
 outside the spot. The colored curves represent different constant magnetic gradients.
 }
 \label{fig:balance}
 \end{figure}

To study the pressure balance in more detail, I take a model for the outer atmosphere 
calculated with the CO$^5$BOLD-code \citep{co5bold}.  
Properties of this model are displayed in \cite{Kupka09}. 
I consider the 
central umbra where curvature forces can be neglected because here magnetic field lines are straight.
Models M3 and M4 of
\cite{SandW75} and \cite{Ketal80} are combined, and the internal gas pressure from this models 
was taken. I assume that the 
magnetic field strength is 3000\,G at the zero-height of the M4-model. With this assumption, the two height 
scales can be adjusted. For each 
height the magnetic field strength to balance the outer gas pressure is calculated, and the 
curve with the symbols in Figure~\ref{fig:balance} is obtained. Linear curves for different 
magnetic gradients are also calculated to enable the comparison. Best agreement is obtained with a gradient 
of $-4$\gkm{} in the lower photosphere. The curve flattens towards higher layers, but the gradient for the upper 
photosphere is still steeper than that for the divergence-free condition.
A very similar picture is obtained when the umbral model of \cite{mackkl} is used instead of the M3/M4 models.

\section{Numerical simulations}\label{S:numsim}

Nowadays, sophisticated numerical simulations of sunspots are available. 
A series of such simulations are presented by \cite{Rempel11a}, \cite{Rempel11b}, 
\cite{Rempel11c} and \cite{Rempel12}. Although the information is contained in the models,
he did not investigate in detail the height dependence of the magnetic field.
A first attempt to determine
the magnetic gradient was done by \cite{Jaeggli12b} who used
the model of \cite{Rempel11c}. They found a gradient of $-1$\gkm{} from the model.

%
 \begin{figure} 
 \centerline{\includegraphics[width=0.98\textwidth, clip=]{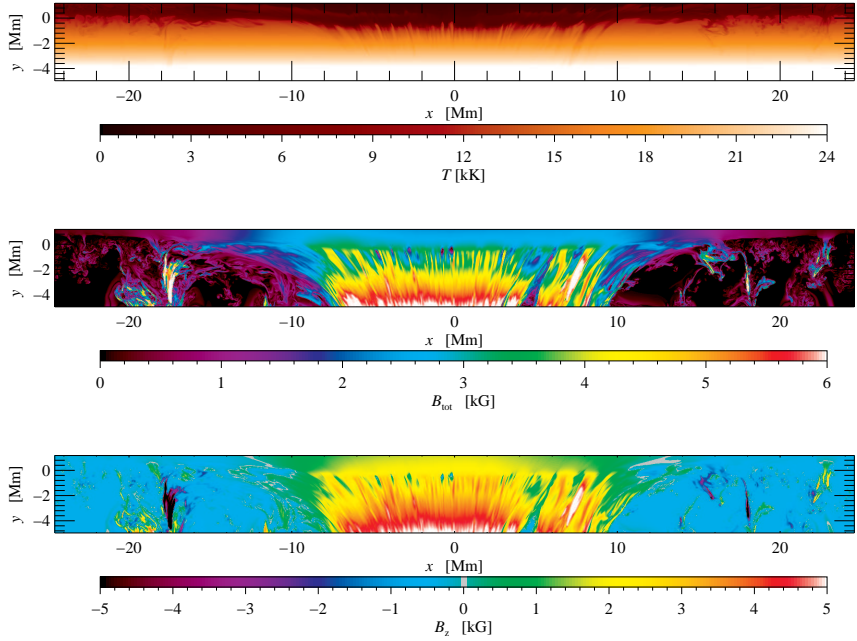}}
 \caption{Cuts through numerical simulations of a sunspot by \cite{Rempel12}. Upper panel: 
 temperature, mid: total magnetic field strength, bottom: vertical magnetic field component.
 }
 \label{fig:rempel}
 \end{figure}

%
 \begin{figure} 
 \centerline{\includegraphics[width=0.98\textwidth, clip=]{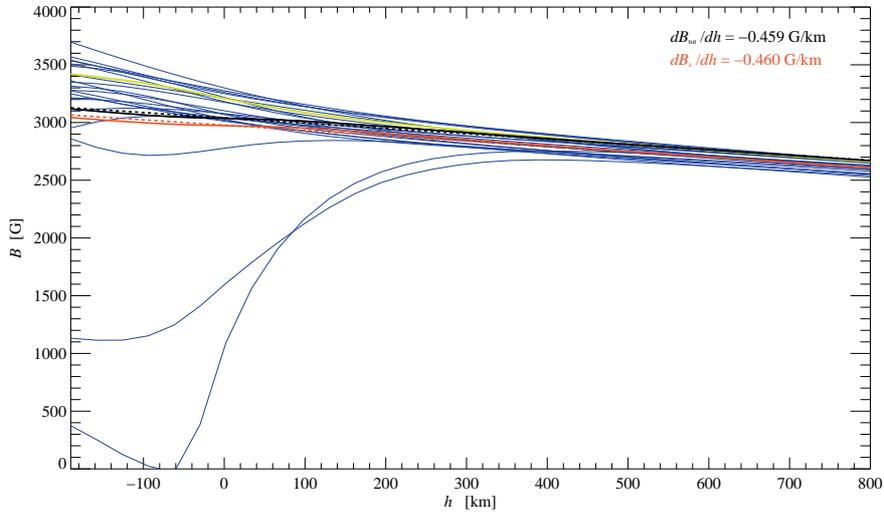}}
 \caption{Magnetic field stratification in the simulation of \cite{Rempel12} for different locations.
          The black curve gives the mean for {$B_{\rm tot}$}{} and the red curve represents {$B_{\rm z}$}. 
          Dashed curves
          display linear fits. The green curve is the mean for {$B_{\rm z}$}{} in a selected small range. 
 }
 \label{fig:gradrempel}
 \end{figure}

For a more detailed investigation, I used the model published in
\cite{Rempel12}. Vertical cross sections of the sunspot for temperature, total magnetic field strength 
and its vertical component are displayed in Figure\,\ref{fig:rempel}.
Obviously the magnetic stratification depends on the location.
At some positions, deep layers are even free of magnetic field. The geometrical height is set to zero 
where lg $\tau$ is zero in the umbra corresponding to a temperature of 4000\,K.
Therefore, Figure~\ref{fig:gradrempel} shows the height dependence for several locations together with mean 
curves for the total magnetic field strength and for the vertical component for a more homogeneous central 
part of the umbra ($-3.78$ to +2.59\,Mm in Figure~\ref{fig:rempel}). Local differences occur mainly 
in the deep layers. The mean curves for {$B_{\rm tot}$}{} and {$B_{\rm z}$}{} differ in the absolute values, 
but not in the gradients. Linear fits to the curves yield 
gradients of $-0.459$\gkm{} and $-0.460$\gkm, respectively. These values are lower than those of 
\cite{Jaeggli12b} and correspond more to those obtained with the divergence-free method. \cite{Rempel12} 
configured the simulation such that the magnetic field strength in the spot drops from 6400\,G at the 
bottom of his calculations to 2560\,G at the top, and the calculation covers a height range of 6144\,km, 
corresponding to a gradient of $-0.625$\gkm.  

The averaged curves show an almost linear magnetic field gradient in all heights, and all curves for single
locations show a linear behavior above a height of 300\,km. For a majority of the curves from single locations,
we see slightly steeper gradients in deep layers, but for a few locations, 
the magnetic field decreases towards 
deep layers, reaching even zero in one case. These locations have a strong impact when taking averages.
The mean curve for the central part [$-0.10$,\,+0.54]\,Mm; green 
curve in Figure~\ref{fig:gradrempel}, where no low field intrusions from below occur, shows a gradient of 
about $-1$\gkm{} for heights below 200\,km. This value 
corresponds to the value found by \cite{Jaeggli12b} and is in less agreement with the results obtained with 
divergence-free method, but it is also lower than the gradient from different photospheric lines.

\section{Discussion}\label{S:discussion}

The discrepancy between the magnetic gradients obtained from different methods was topic of the workshop
``The problem of the magnetic field gradients in sunspots''
held in Meudon October 3 -- 5, 2016 (in the following called the Meudon workshop). In this section, I pick up 
many aspects of the discussion at the workshop.

Maxwell's theory of electromagnetism is well established, thus there are no doubts that the 
magnetic field is divergence-free. Local problems can arise when the spatial resolution is 
insufficient, and neighboring pixels belong to different fine structures which might reflect 
even different height layers. Figures\,\ref{fig:BandG_bbh} and \ref{fig:tiwari8} demonstrate the 
internal structure of sunspots. Such a structure would create a bias of the horizontal partial derivatives,
but it is hard to imagine how this effect could cause a systematically reduced decrease of the 
vertical magnetic field strength. It should also cause an increase at another location, and on 
average over the whole spot, it should show large fluctuations, but no systematic bias of the 
magnetic gradient. A more sophisticated investigation of this problem, suggested by Jean Heyvaerts,
was discussed at the Meudon workshop leading to the result that this effect does not cause a significant 
violation of div\,${\bb}$ = 0. 
For more details see \cite{Bommier14}. 
At the Meudon workshop, Silvano Bonazzola suggested to investigate the effect of applying quotients of 
finite differences as approximations 
instead of the real derivatives to determine div ${\bb}$. If the spatial variation scale 
of magnetic field is large with respect to the pixel size, finite difference and 
derivative are not very different, whereas if the magnetic field spatial variation scale is small 
or comparable to the pixel size, a difference may exist. But in this case, if the field lines are 
distorted inside a pixel face, the magnetic flux goes in and out as well, and the flux remains 
balanced. In this case, it remains unclear how the field line distortion can be responsible for an apparent 
flux loss, such as the one observed with a factor of about 10 between vertical and horizontal flux.
In other words, the difference between finite difference and derivative results are too small 
to explain the observed difference between the gradients.

On the other hand, for a small sunspot the magnetic field strength reduces from an umbral value of 
about 2500\,G to about 500\,G at the outer edge of the penumbra. The radius of such a spot is about 
10\,000\,km, and the horizontal gradient is in the order of 0.2\gkm, too. And a steep gradient of $-2$\gkm{}
can be valid only over a limited height range. A magnetic field strength of 3000\,G in deep layers would 
be reduced to zero at a layer 1500\,km higher.

\subsection{Formation height of spectral lines}

Another uncertainty is the formation height of the spectral lines. Spectral lines are formed in 
an extended height range, and the mean value or the center of gravity of the contribution function 
or the response function are given as formation height. Nevertheless, height differences of 100\,km
or even more are used to obtain magnetic height gradients of $-2$\gkm{} from the height difference method. 
To reduce these gradients, one has to increase the height differences by a factor of 5 -- 10. 
The theory of stellar atmospheres tells us that the solar photosphere has an extension of less than 
500\,km, and with modern solar telescopes we resolve such a range, and we can prove that the extension 
is not larger than 500\,km. \cite{faurobert09} directly measured the height difference 
of the two iron lines at 630.15\,nm and 630.25\,nm and found a value of 64\,km, almost independent of the 
location on the Sun. This is the same difference as it was used by \cite{Betal13} for their {\it Hinode} data.
The uncertainty of the formation height is not the way out of the problem of the magnetic gradients.

In general, the linear polarization is weaker than the circular one. Therefore, the response of linear 
polarization to the magnetic field happens in deeper layers than that of circular polarization. This 
may impact the results for the magnetic vector, but a sophisticated inversion code is able 
to take this effect into account, if the observations contain enough height dependent information, and the 
physical parameters are calculated height dependent. Nevertheless, such inversions still yield steep 
magnetic gradients from height differences.

One may also argue that the results are affected from the properties of different detectors and 
different telescopes. However, \cite{Cuberes05} and \cite{BandG08} obtained their data from two spectral 
lines strictly simultaneously on the same detector, and both teams got steep gradients, although 
\cite{Cuberes05} performed a height dependent inversion, while \cite{BandG08} inverted their data 
independently for the two lines.

\subsection{Does return flux play a r{\^ o}le?}

%
 \begin{figure} 
 \centerline{\includegraphics[width=0.98\textwidth,clip=]{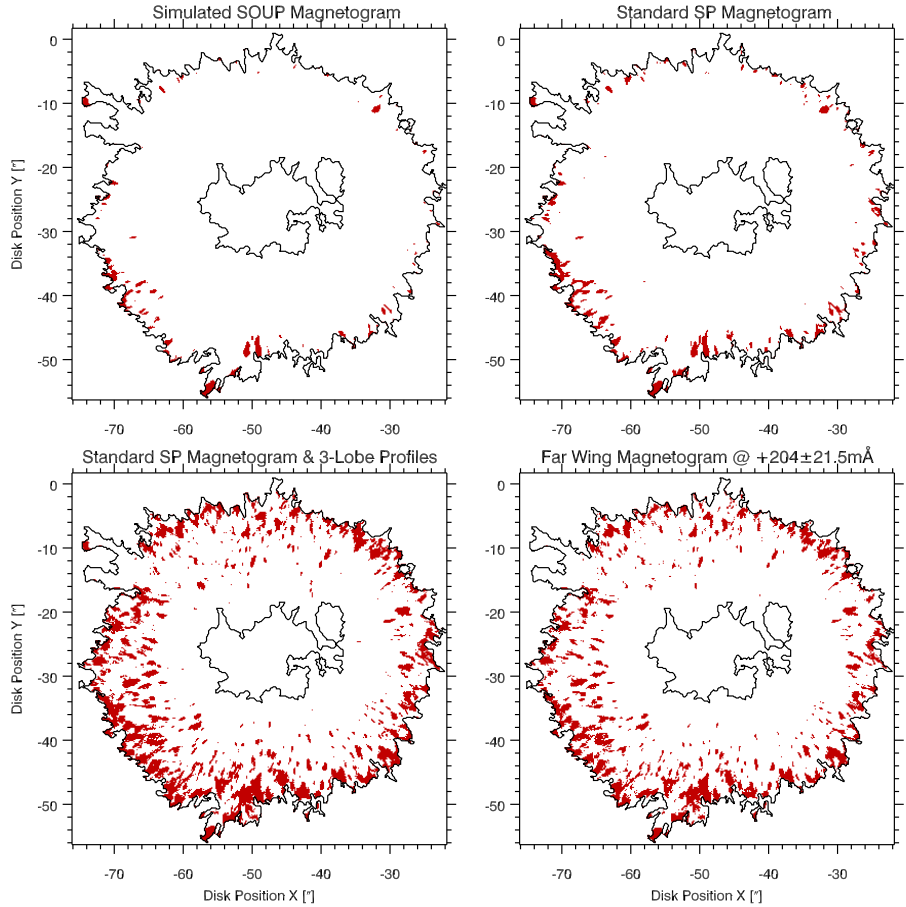}}
 \caption{Distribution opposite polarities in the penumbra. Top left: from simulated SOUP/SST magnetogram, 
       top right: from standard {\it Hinode}-SP magnetogram, bottom left: from a magnetogram taking 3-lobe 
       profiles into account, and bottom right: from a magnetogram obtained in the far wings of the line profile.
       Credit: Franz \& Schlichenmaier, 
       2013, reproduced with permission \copyright ESO.
}
 \label{fig:franz6}
 \end{figure}

There are several hints that a part of the magnetic field lines in 
the outer penumbra are curved downward \citep{Westendorp01, delToro01, BBC04, RandA12, Fetal16}. 
\cite{OandG89} developed a sunspot model with return flux in the outer penumbra and found a good agreement 
with observations of unipolar and approximately round sunspots.
In a detailed observational 
study, \cite{FandS13} compared the occurrence of opposite polarities from the standard {\it Hinode}-SP magnetogram, 
from a simulation of 
a magnetogram mimicking observations with the Lockheed \textit{Solar Optical Universal Polarimeter} (SOUP; 
\citealp{SOUP}) at 
the \textit{Swedish Solar Tower} (SST; \citealp{SST}), from a {\it Hinode}-SP magnetogram calculated for the 
far wings of the spectral line, and from a magnetogram considering 3-lobe profiles. While for standard magnetograms 
such as the first two only 4\% of the penumbral pixel exhibit opposite polarities, the number increases to
about 17\% for the latter, more sensitive methods. \cite{FandS13} conclude that opposite magnetic polarities 
are more frequent than one previously thought based on low spectral resolution measurements.
The distribution of opposite polarities for the different types of magnetograms is displayed in 
Figure~\ref{fig:franz6}.

Let us assume the spot consists of a bundle of flux tubes rising through the umbra. According to this amount 
of magnetic flux and the pressure conditions, the magnetic field configuration is adjusted with steep 
vertical and horizontal gradients. However, the horizontal field in the penumbra is extended, and therefore
additional rising flux is required. Such flux may occur as almost horizontal flux tubes independent 
from the main spot.
If flux rises and  descends close together inside the penumbra, the additional flux will not 
be recognized at limited spatial resolution. Unresolved, it will look like an extended horizontal magnetic field. 
If resolved, one will see a salt-and-pepper pattern of ascending and descending magnetic field lines,
and a decrease and increase of the horizontal components of the magnetic field.
Then the horizontal gradient of the magnetic field 
appears shallower than in vertical direction above the umbra. The horizontal decrease according to the umbral flux
will correspond to the vertical 
decrease, but new magnetic flux replenishes the horizontal field in lower layers of the outer penumbra. 
For the penumbra, this effect may be an explanation for the gradient differences among the different methods, 
but is hard to imagine that 
such an effect also acts in the umbra, where the magnetic field is less inclined.

\subsection{Twist and writhe}

Simple models for sunspots with twisted magnetic field have been presented by \cite{OandF83} who 
considered a twist of the whole sunspot. The twist in these models is smaller in higher atmospheric layers and
influences the structure of the sunspot only slightly. \cite{Seehafer90} studied the current helicity 
in sunspots and found the same sign of helicity in one hemisphere and the opposite sign in the other.
This scenario was confirmed by \cite{Pevtsov95} and \cite{Zhang02}. However, it is a tendency but not a 
strict rule.
As shown in Figure\,\ref{fig:BandG_bbh}, the magnetic field exhibits a local structure, which may arise 
from twist and writhe of bundles of flux 
tubes. As a consequence, the magnetic azimuth will rotate over a certain height range, and part of the 
polarization of the light will be canceled out. In this case, one may obtain a reduced magnetic field 
strength from 
spectral lines formed over a wider height range, which probe also the higher layers of the photosphere.
A spectral line formed in deep layers probes only a narrow atmospheric layer and exhibits a strong magnetic field.
There have been indications that magnetic twist plays a  r{\^o}le in sunspots, especially in the penumbra.
\cite{Borrero08} found opposite signs for the deviation of the azimuth from the radial direction on both 
sides of penumbral filaments, and explained this result by a magnetic field wrapping around the filament.
An analysis of the torsion within a sunspot was performed by \cite{Socas05} who concluded 
that flux ropes of opposite helicity may coexist in the same spot.

\begin{table}
 \caption{Magnetic field strengths resulting from inversions with twisted magnetic field in the provided line 
         profiles }
  \label{tbl:twist}
 \begin{tabular}{lcccc}     
 \hline
  Model   &  Si\,{\sc i}\,1082.7\,nm & Ca\,{\sc i}\,1083.9\,nm &    Fe\,{\sc i}\,630.25\,nm & 
  Fe\,{\sc i}\,630.15\,nm  \\
  \hline
  M4\_30\_00\_0  & 2650\,G      & 2740\,G  &   2710\,G  &   2720\,G    \\
  M4\_05\_00\_0  & 2930\,G      & 2940\,G  &   2950\,G  &   2950\,G    \\
  M4\_05\_10\_1  & 2900\,G      & 2930\,G  &   2930\,G  &   2910\,G    \\
  M4\_05\_20\_1  & 2830\,G      & 2920\,G  &   2850\,G  &   2760\,G    \\
 \hline
 \end{tabular}
 \end{table}
  
%
 \begin{figure} 
 \centerline{\includegraphics[width=0.98\textwidth,clip=]{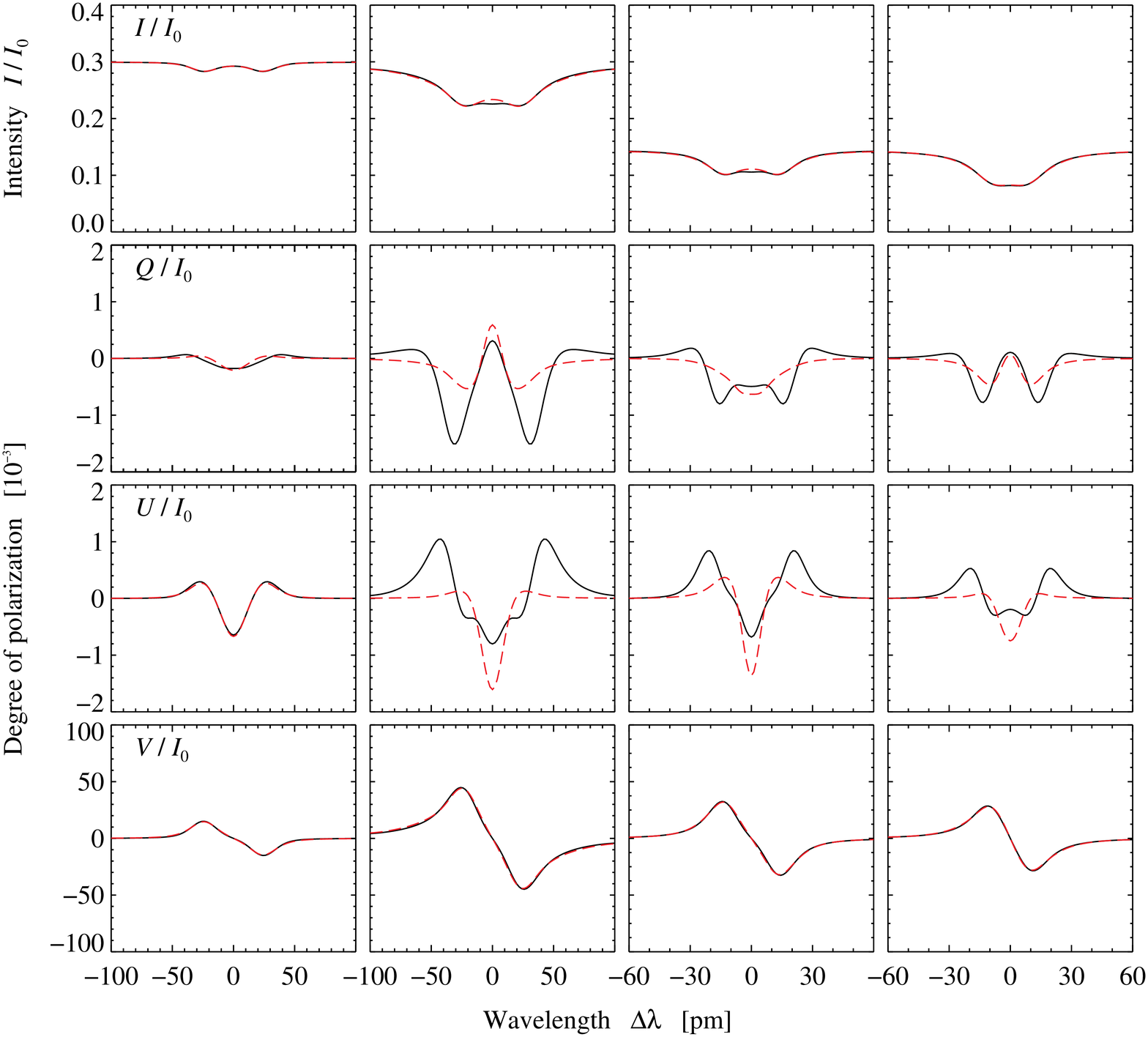}}
 \caption{Comparison of model profiles (black lines) with the results of their inversions (red dashed lines).
         From top to bottom: Stokes-$I$, $Q$, $U$, and $V$. 
         Columns from left to right: Ca\,{\sc i}\,1083.9\,nm, Si\,{\sc i}\,1082.7\,nm, Fe\,{\sc i}\,630.25\,nm
         and Fe\,{\sc i}\,630.15\,nm.
 }
 \label{fig:twistprofiles}
 \end{figure}

To investigate this idea, a test is performed. The umbral model atmosphere M4 of \cite{Ketal80} 
with a height dependent magnetic field strength, but with zero values for the magnetic azimuth and 
inclination serves 
as starting model. The height range of the model is extrapolated up to lg$\tau = -5.4$.
In the starting model M4\_30\_00\_0, the magnetic field decreases from 3000\,G at lg$\tau$ = 0 
to 1500\,G at lg$\tau = -5.0$, corresponding to a magnetic gradient of about $-3$\gkm.
In the next step, this model is modified that the magnetic field strength decreases only 
to 2750\,G at lg$\tau = -5.0$,
corresponding to a magnetic gradient of about $-0.5$\gkm{} (M4\_05\_00\_0).
Then an azimuth rotating with height and a height independent inclination are introduced. 
I assume one rotation over five units of lg~$\tau$. The inclination is 10$^\circ$ (M4\_05\_10\_1)
and 20$^\circ$ (M4\_05\_20\_1).
Thus the inclination points into different directions according to the magnetic azimuth,
and therefore a magnetic field is modeled that winds around the line-of-sight. 
The corresponding Stokes-profiles are calculated for the spectral lines Si\,{\sc i}\,1082.7\,nm,  
Ca\,{\sc i}\,1083.9\,nm, Fe\,{\sc i}\,630.15\,nm and Fe\,{\sc i}\,630.25\,nm. These lines in general probe 
rather different 
layers of the photosphere, the Ca line forms in deep layers, while the Si line probes the upper photosphere 
(see \citealp {Betal16}, \citealp{Felipe16}). However, in the cold umbra, the height difference between the 
calcium line
and the silicon line reduces to about 50\,km. To calculate the line profiles, I used a macroturbulence of 
2.0\,km\,s$^{-1}$, leading to a broadening of the profiles.
Subsequently, SIR-inversions are performed with the line profiles from the different models
independently for the individual lines. In these inversions, the magnetic parameters are treated as 
height-independent. This way, one obtains an average of the magnetic azimuth integrated over a 
certain height range weighted 
with the response or contribution function. The obtained linear polarization depends on this. 
The Ca line exhibits only a small change of the magnetic field strength with increasing number of 
azimuth windings, but the field strength from the silicon line decreases rapidly as Table\,\ref{tbl:twist} shows.  
The obtained gradients of the magnetic field are a bit shallower than the ones preset in the starting model, 
probably an effect of the 
introduced broadening. For M4\_30\_00\_0 the gradient is $-2$\gkm and for M4\_05\_00\_0 it is $-0.2$\gkm. 
Including twist, the gradients become steeper, $-0.7$\gkm{} for an inclination of 10$^\circ$ 
and $-1.5$ -- $-2$\gkm{} for an inclination of 20$^\circ$.
A problem arises because Stokes-$Q$ and $U$ then have many lobes of varying sign, what is not observed in sunspots. 
A comparison of the created profiles and the fits is given in Figure\,\ref{fig:twistprofiles}. Note that with the 
settings for the inversions, the linear polarization cannot be reproduced exactly. 
Perhaps this point is less important because the linear polarization created in the models is rather low.

In conclusion, twist and writhe may play a r{\^ o}le, but the solar reality is not 
as simple as these tests. Another problem of such a scenario may be that many windings lead to magnetic 
instabilities such as the kink instability (see \citealp{kinkinst}).
It also remains unclear if this simple configuration fulfills the condition div~$\bb$~=~0.
More detailed and sophisticated investigations are required to arrive at a conclusive result about
how twist and writhe can contribute to solve the discrepancy of the magnetic gradients from different methods.

\subsection{Other ideas to explain the discrepancy of magnetic gradients from different methods}

\cite{Bommier13} discussed that the Debye-volume is not a sphere because of the stratification of 
the solar atmosphere, and as a consequence typical horizontal and vertical length scales are different. 
This requires a re-scaling of typical lengths and affects the partial derivatives of the magnetic field,
especially the vertical one. Bommier considered the vertical pressure scale height in comparison to the 
diameter of a typical photospheric structure such as a granule. However, the granular velocity structure can 
be followed over several hundred kilometers (almost the whole photospheric extension), and thus the aspect ratio
is much smaller than she needs to explain the magnetic gradients. More relevant are the velocities.
Over this height range, the vertical velocity changes from about 2\,km\,s$^{-1}$ to a horizontal 
expansion of about 1\,km\,s$^{-1}$ \citep{B85}. This velocity pattern characterizes a granule. 
Again the ratio is too small. The situation is similar for sunspots which can be traced down to 40\,Mm 
as shown by \cite{Chen98} using helioseismic investigations. The discussion is ongoing,
and perhaps the velocity difference between ions and electrons plays a more important r{\^ o}le.

If a sunspot is composed of many flux tubes, the filling factor for the magnetic field may vary 
in different heights. This idea was suggested by Gerard Belmont 
and discussed during the Meudon workshop.
A consideration by Pascal D\'emoulin at this workshop showed 
that a filling factor smaller than one will make the situation even worse, 
\textit{i.e.} such a scenario will create steeper gradients.

Refraction in the solar atmosphere was suggested as a possible source of errors, but several 
investigations showed that refraction is negligible in the solar photosphere and chromosphere. 
\cite{Vincent13} provide an equation for the refractive index including relativistic effects 
(their Equation 13).
If we take the hydrogen density from model C of \cite{Vernazza76} at lg $\tau$ = 0, the refractive index 
amounts to 1 + 5$\times 10^{-7}$. Thus the deviation from unity is very small, and effects due to refraction 
can be neglected in the solar atmosphere. Even if they are not negligible, they will 
occur when observing perpendicular to the stratification of the atmosphere near the limb, 
while most sunspots were observed close to disk center, {\it i.e.} more along the density gradient. 
Therefore, atmospheric refraction is unlikely to contribute to the problem of the magnetic gradients.

\section{Conclusions}

The problem of the magnetic gradient in the photosphere of sunspots and pores remains unsolved.
Observed gradients are much steeper when they are determined from different spectral lines or from
single line making use of the atmospheric stratification than when the gradients are determined 
from the condition {div\,$\bb$\,=\,0}. In most cases, from the first method values between $-2$\gkm{} and 
$-3$\gkm{} were obtained, while applying the divergence-free condition yields values of $-0.5$\gkm{} or 
even shallower. Additional investigations were presented, partly supporting steep gradients, {\it e.g.}
the pressure balance to the stratification of the outer surroundings, and partly shallow gradients,
{\it e.g.} the determination from a numerical simulation. 
All methods based on information from chromosphere lines exhibit shallow gradients
of $-0.5$\gkm{} or shallower. Radio observations cover an even larger height range and yield shallow
gradients between $-0.1$\gkm{} and $-1.4$\gkm. 
Since we know that the gas pressure of the quiet Sun decays exponentially with height, one should also expect
a shallow gradient of the magnetic field for chromospheric layers and above.
Thus the discrepancy between the different methods 
appears mainly in the photosphere. 

Several options to explain the difference between the methods were discussed in the previous 
section, but most of them can be ruled out. 
The explanation cannot be found in the uncertainties of the height scale determinations nor 
in the influence of spatial resolution in the determination of the partial derivatives of the magnetic field.
It appears that the final word is not spoken for
the effects of twist and writhe. It seems likely that here at least a part of the answer 
could be found, but more sophisticated investigations are required, supported 
by numerical simulations.

Observations with the upcoming new solar facilities are also required. These instruments provide
a better spatial resolution and will allow us to study the fine structure in sunspots in more 
details. There is indication that sunspots have an internal structure, not only when umbral dots appear, 
but also in the normal dark umbra. For example, with such high resolution investigations, it should be 
possible to obtain a clearer picture about twisted magnetic fields than we have at the moment, 
and judge how much twist and writhe
really contribute to the problem of the magnetic gradients. However, better observations with higher 
resolution and polarimetric accuracy and sophisticated numerical simulations may reveal other 
explanations.

%
 \begin{acks}
   I am deeply indebted to Dr. V{\'e}ronique Bommier for many discussions and comments on the topic. 
   I also thank her and Prof. Carsten Denker for carefully reading the manuscript.
   Dr. Matthias Steffen provided me with a model atmosphere
   and Dr. Matthias Rempel with a cut through one of his numerical simulations.
   My thanks go also to Dr. Morten Franz and Dr. Sanjiv Tiwari for the permission to use 
   their figures (Figure\,\ref{fig:franz6} and Figure\,\ref{fig:tiwari8}).  
   The 1.5-meter GREGOR solar telescope was built by a German 
consortium under the leadership of the Kiepenheuer Institute for Solar Physics 
in Freiburg with the Leibniz Institute for Astrophysics Potsdam, the Institute for Astrophysics 
G\"ottingen, and the Max-Planck-Institute for 
Solar System Research in G\"ottingen as partners, and with contributions by the Instituto de 
Astrof{\'\i}sica de Canarias and the Astronomical 
Institute of the Academy of Sciences of the Czech Republic.  
 \end{acks}

\vspace{5mm}

\noindent
\textbf{Disclosure of Potential Conflicts of Interest}\quad The author declares that he has no 
conflicts of interest.

%
%
\bibliography{spot_grad.bbl}  
%

\end{article} 
\end{document}